\documentclass[aps,superscriptaddress,twocolumn,twoside,floatfix,pra,nofootinbib,a4paper]{revtex4-1}
\usepackage{graphicx,amsmath,amssymb,color}
\usepackage[normalem]{ulem}
\usepackage{amsfonts}
\usepackage[ocgcolorlinks,colorlinks=true,linkcolor=blue,citecolor=red]{hyperref}
\usepackage{latexsym}
\usepackage{amsfonts}
\usepackage{amsthm}
\usepackage{mathrsfs}
\usepackage{natbib}
\usepackage{color,verbatim}
\usepackage{psfrag}
\DeclareMathAlphabet{\mathrsfs}{U}{rsfs}{m}{n}
\DeclareMathAlphabet{\mathpzc}{OT1}{pzc}{m}{it}
\DeclareMathAlphabet{\matheus}{U}{eus}{m}{n}
\DeclareMathAlphabet{\mathbbold}{U}{bbold}{m}{n}

\newcommand{\ba}{\begin{eqnarray}}
\newcommand{\be}{\begin{equation}}
\newcommand{\ee}{\end{equation}}

\newcommand{\ea}{\end{eqnarray}}
\newcommand{\ban}{\begin{eqnarray*}}
\newcommand{\ean}{\end{eqnarray*}}
\newcommand{\Tr}{\operatorname{tr}}

\newcommand{\kets}[2]{|#1\rangle_{\!_{#2}}}
\newcommand{\ket}[1]{|#1\rangle}
\newcommand{\bra}[1]{\langle#1|}

\newcommand{\expect}[1]{\langle#1\rangle}

\newcommand{\ie}{{\it{i.e.}~}}
\newcommand{\etal}{{\it{et al.}}}
\newcommand{\rma}{\mathrm{a}}
\newcommand{\rmb}{\mathrm{b}}

\begin{document}

\title{Nonlocal correlations in the star-network configuration}

\author{Armin Tavakoli}
\affiliation{Department of Physics, Stockholm University, S-10691 Stockholm, Sweden}
\affiliation{ICFO-Institut de Ciencies Fotoniques, Mediterranean Technology Park, 08860 Castelldefels (Barcelona), Spain}

\author{Paul Skrzypczyk}
\affiliation{ICFO-Institut de Ciencies Fotoniques, Mediterranean Technology Park, 08860 Castelldefels (Barcelona), Spain}

\author{Daniel Cavalcanti}
\affiliation{ICFO-Institut de Ciencies Fotoniques, Mediterranean Technology Park, 08860 Castelldefels (Barcelona), Spain}

\author{Antonio Ac\'in}
\affiliation{ICFO-Institut de Ciencies Fotoniques, Mediterranean Technology Park, 08860 Castelldefels (Barcelona), Spain}
\affiliation{ICREA--Instituci\'o Catalana de Recerca i Estudis Avan\c{c}ats, Lluis Companys 23, 08010 Barcelona, Spain}

\begin{abstract}
The concept of bilocality was introduced to study the correlations which arise in an entanglement swapping scenario, where one has two sources which can naturally taken to be independent. This additional constraint leads to stricter requirements than simply imposing locality, in the form of bilocality inequalities. In this work we consider a natural generalisation of the bilocality scenario, namely the star-network consisting of a single central party surrounded by $n$ edge parties, each of which shares an independent source with the centre. We derive new inequalities which are satisfied by all local correlations in this scenario, for the cases when the central party performs (i) two dichotomic measurements (ii) a single Bell state measurement. We demonstrate quantum violations of these inequalities and study both the robustness to noise and to losses. 
\end{abstract}

\maketitle

\section{Introduction}

Bell's theorem imposes constraints on the correlations between space-like separated events that have a common source \cite{Bell64,BCP+13}. The typical Bell scenario consists of two distant parties, Alice and Bob, who apply respective measurements $x$ and $y$ with possible outcomes $a$ and $b$ on a bipartite physical system. The Bell locality assumption consists of considering that the possible correlations observed between the outcomes of Alice and Bob's measurements are due to a set of common past factors, denoted by $\lambda$, unknown to them. In mathematical words, this amounts to saying that the probability distribution of the outcomes $a$ and $b$ given their measurement choices $x$ and $y$ is given by 
\ba\label{local p}
P(ab|xy)=\int q(\lambda)P(a|x,\lambda)P(b|y,\lambda)d\lambda,
\ea
where $q(\lambda)$ is a probability distribution over the set of possible past factors. It turns out that local measurements on entangled states can generate some probability distributions that are not of this form. For this reason it is said that these correlations are nonlocal. 

Another remarkable feature of quantum systems is the fact that systems that have never interacted can become nonlocally correlated. This can happen through the process of entanglement swapping \cite{ZZHA93}, in which two independent pairs of entangled particles are first created and then one particle from each pair is jointly measured. As a result the other particles become entangled. 

It is natural to expect that the correlations obtained in an entanglement swapping experiment should be even more difficult to simulate with a local theory. In order to formalise this intuition, Branciard \etal~ proposed the concept of bilocality \cite{BGP10,BraRosGis12} in the following scenario: a source $S_1$ sends particles to two distant observers Alice and Bob. Independently, another source $S_2$ sends particles to Bob and another observer Charlie. Alice, Bob and Charlie perform measurements labelled by $x, y$ and $z$ with outcomes $a, b$ and $c$ respectively. The bilocal assumption consists in considering that, since the sources $S_1$ and $S_2$ are independent, the correlations observed by Alice, Bob and Charlie are given by independent causes $\lambda_1$ and $\lambda_2$ in the following way:

\begin{multline}\label{bilocal p}
P(abc|xyz)=\iint q_1(\lambda_1)q_2(\lambda_2)P(a|x,\lambda_1)\\
\times P(b|y,\lambda_1\lambda_2)p(c|z,\lambda_2)d\lambda_1d\lambda_2.
\end{multline}
The key point of this decomposition is that the results of Alice's (Charlie's) measurements are only determined by the measurement input $x$ ($z$) and the variable $\lambda_1$ ($\lambda_2$). Moreover, the variables $\lambda_1$ and $\lambda_2$ are independent. 

There are several motivations for studying these new nonlocality scenarios. First, from a fundamental point of view, it is important to determine natural classical models that can not explain the predictions of quantum mechanics to understand the strength of quantum correlations. From an applied point of view, nonlocal correlations have been seen as a resource for device-independent protocols (see e.g. \cite{BCMW10,ABGM+07,PAM+10,RHC+11}), so it is important to establish from which scenarios it is possible to extract nonlocal correlations. Finally, long-distance quantum networks are the main goal of quantum communication, so it is fundamental to study their nonlocal capabilities.

In the present paper we generalise the idea of bilocality to the star-network configuration (see Fig.~\ref{f: setup}). We develop inequalities satisfied by $n$-local models and show that they are violated by correlations obtained in appropriate multipartite entanglement swapping experiments. We also discuss the robustness to noise of the initial pairs of particles that are needed in order to violate the obtained inequalities and show that it is independent of the number of nodes in the star-network.

\section{$n$-locality in the star-network scenario}

\begin{figure}
\centering
\includegraphics[width=0.9\columnwidth]{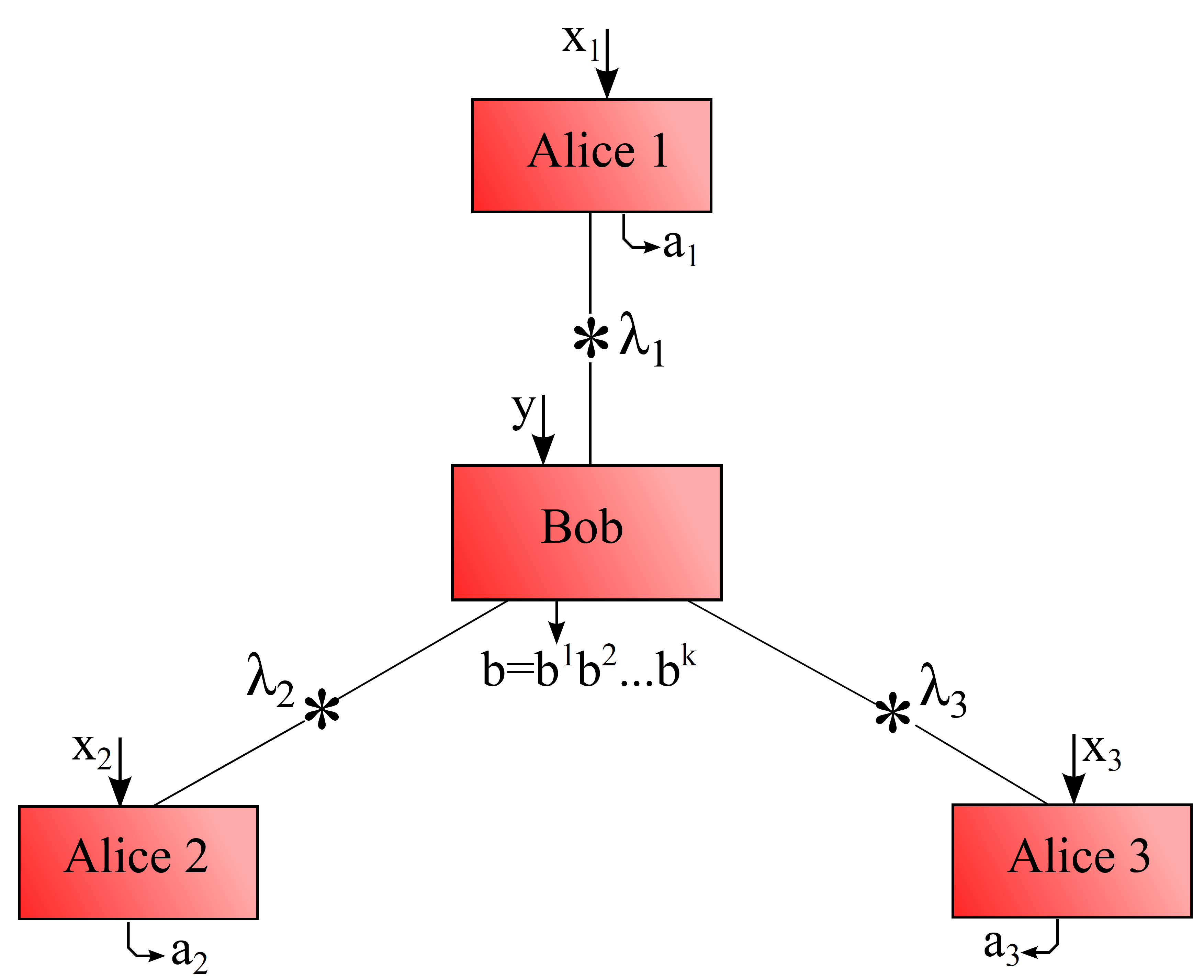}
\caption{Star-network measurement scenario under the 3-locality assumption.}\label{f: setup}
\end{figure}

In what follows we shall consider the star-network composed by $n+1$ parties (see Fig.~\ref{f: setup}), where a central node (referred to as Bob) shares an entangled state with each of  $n$ nodes (referred to as the Alices). Here we will restrict our attention to the special case where each of the $n$ Alices performs $m_\rma = 2$ measurements each with $d_\rma = 2$ possible outcomes. Each outer party is referred to as party $i$, for $i\in \mathbb{N}_{n}$. The measurements performed by party $i$ are labelled by $x_i\in\{0,1\}$ and the corresponding outcomes denoted $a_i\in\{0,1\}$. The central node will perform one out of $m_\rmb$ measurements labelled by $y\in\{0,1,\ldots,m_\rmb-1\}$ with  $d_\rmb = 2^{k}$ outcomes labelled by a string $b=b^1b^2\ldots b^k$ where $b^i\in\{0,1\}$ for $i=1,2,\ldots,k$. We shall refer to the scenario where Bob performs $m_\rmb$ measurements and has a $k$-bit outcome as the ``Bob $m_\rmb \to 2^k$ Scenario". 

We can now define the concept of $n$-locality in the star-network configuration. Similarly to the bilocal case \eqref{bilocal p} we say that a probability distribution on the star-network scenario is $n$-local if it can be written as
\begin{multline}\label{nlocal p}
P(a_1,\ldots,a_{n},b|x_1,\ldots,x_{n},y)=\\ \int  \left(\prod_{i=1}^{n} d\lambda_i q_i(\lambda_i)P(a_i|x_i,\lambda_i)\right) P(b|y,\lambda_1,\ldots,\lambda_n). 
\end{multline}
In what follows we will develop inequalities that are satisfied by all probability distributions of the form \eqref{nlocal p}, but which can be violated measuring quantum states distributed in a star-network configuration. These inequalities are the analogues of Bell inequalities in the star-network locality scenario.  

\subsection{Bob $2\rightarrow 2$ Scenario}
In this section we consider the star-network scenario when Bob, at the central node, can choose two possible measurements with two possible outcomes each, i.e. $k = 1$. This is thus the Bob $2\rightarrow 2$ Scenario. This scenario is motivated in part by experimental limitations, where, for instance, partial Bell state measurements are more feasible than complete Bell state measurements.

We start by introducing the following correlation functions for the $n+1$ party measurement outcomes
\begin{multline}\label{n correlator}
\langle  A_{x_1}^1 \cdots A_{x_n}^n B_y \rangle \equiv \\ 
\sum_{a_1\cdots a_n b} (-1)^{b+\sum_i a_i} P(a_1\cdots a_nb|x_1\cdots x_ny),
\end{multline}
and the following functions:
\begin{equation}\label{I}
I=\frac{1}{2^n} \sum_{x_1\cdots x_n} \langle A_{x_1}^1 \cdots A_{x_n}^n B_0 \rangle. \newline
\end{equation}
\begin{equation}\label{J}
J=\frac{1}{2^n} \sum_{x_1\cdots x_n} (-1)^{\sum_i x_i} \langle A_{x_1}^1 \cdots A_{x_n}^n B_1 \rangle .
\end{equation}
Now our first result can be stated:  
\begin{flushleft}
\textbf{Theorem 1 (Bob $2 \rightarrow 2$ $n$-locality)}
\end{flushleft}
Consider the star-network scenario composed by $n+1$ parties applying two dichotomic measurements, \ie $m_\rma = d_\rma = m_\rmb = 2$, $k=1$. In this scenario, every $n$-local probability distribution $\eqref{nlocal p}$ satisfies the following inequality
\begin{equation}\label{e:ineq 2 to 2}
\lvert I \rvert ^{1/n}+\lvert J \rvert ^{1/n} \leq 1.
\end{equation}
\textbf{Proof:}\newline\newline
We start with considering only the quantity $I$. From eqs. \eqref{nlocal p}, \eqref{n correlator} and \eqref{I} we have
\begin{multline}
\!\!I = \frac{1}{2^n} \sum_{\substack{x_1\cdots x_n\\a_1\cdots a_nb}}(-1)^{\sum_i a_i + b} \int \left(\prod_{i=1}^n q_i(\lambda_i)P(a_i|x_i \lambda_i)\right)\\ \times P(b|y=0,\lambda)d\lambda,
\end{multline}
where as a shorthand we have written $\lambda = \lambda_1\cdots\lambda_n$ and $d\lambda = d\lambda_1 \cdots d\lambda_n$. 
By re-grouping terms and splitting the sum over $a_1\cdots a_nb$ we find
\begin{multline}
I = \frac{1}{2^n} \sum_{x_1\cdots x_n} \int \left(\prod_{i=1}^n q_i(\lambda_i)\sum_{a_i} (-1)^{a_i}P(a_i|x_i \lambda_i) \right) \\
\times\sum_{b} (-1)^{b}P(b|y=0,\lambda)d\lambda .
\end{multline}
Consider now new correlators constructed from this expression, conditioned on the hidden variables
\begin{align}
\langle A_{x_i}^i \rangle_{\lambda_i} &= \sum_{a_i} (-1)^{a_i} P(a_i|x_i\lambda_i). \\
\langle B_y \rangle_\lambda &= \sum_b (-1)^{b} P(b|y\lambda).
\end{align}
With these new correlators $I$ takes the form
\begin{align}
I &= \frac{1}{2^n} \sum_{x_1,...,x_n} \int \left( \prod_{i=1}^n q_i(\lambda_i)  \langle A_{x_i}^i \rangle_{\lambda_i}\right) \langle B_0 \rangle_{\lambda}d\lambda, \nonumber \\
&=  \frac{1}{2^n}\int\prod_{i=1}^n\left( q_i(\lambda_i)\sum_{x_i}\langle A_{x_i}^i \rangle_{\lambda_i}\right)\langle B_0 \rangle_{\lambda}d\lambda.
\end{align}
By taking the absolute magnitude and noting that $|\langle B_y \rangle_{\lambda}| \leq 1$, after a small amount of re-arranging we arrive at
\begin{equation}\label{e:final I}
|I| \leq \prod_{i=1}^n\left( \frac{1}{2}\int q_i(\lambda_i)\left|\sum_{x_i}\langle A_{x_i}^i \rangle_{\lambda_i}\right|d\lambda_i \right).
\end{equation}
An analogous analysis for the quantity $J$ yields
\begin{equation}
|J| \leq \prod_{i=1}^n\left( \frac{1}{2}\int q_i(\lambda_i)\left|\sum_{x_i}(-1)^{x_i}\langle A_{x_i}^i \rangle_{\lambda_i}\right|d\lambda_i \right).
\end{equation}

Although it is hard to work directly with these two quantities, we can make use of the following lemma:
\newline\newline
\textbf{Lemma 1:} Let $x_i^k$ be non-negative real numbers and $m,n\in\mathbb{N}$, then
\begin{equation}\label{e: lemma 1}
\sum_{k=1}^{m}\left(\prod_{i=1}^{n} x_i^k\right)^{1/n} \leq \prod_{i=1}^{n} \left(x_i^1+x_i^2+...+x_i^m\right)^{1/n}.
\end{equation}
The proof can be found in the Appendix.  Using this lemma we arrive at the following inequality 
\begin{multline}\label{e: after lemma 1}
\lvert I \rvert^{1/n} + \lvert J \rvert^{1/n} \leq \left[\prod_{i=1}^n \frac{1}{2}\int q_i(\lambda_i)\left(\left|\sum_{x_i}\langle A_{x_i}^i \rangle_{\lambda_i}\right| \right.\right. \\ \left.\left.+\left|\sum_{x_i}(-1)^{x_i}\langle A_{x_i}^i \rangle_{\lambda_i}\right|\right)d\lambda_i\right]^{1/n}.
\end{multline}
Using the fact that $x_i$ takes only values $0$ and $1$, an upper bound can easily be seen to hold, namely
\begin{equation}\label{e: upper bound}
\frac{\lvert \langle A_0^i \rangle_{\lambda_i}+\langle A_1^i \rangle_{\lambda_i}\rvert}{2}+\frac{\lvert \langle  A_0^i\rangle_{\lambda_i}-\langle A_1^i\rangle_{\lambda_i}\rvert}{2}\leq 1.
\end{equation}
This yields 
\begin{equation}
\lvert I \rvert^{1/n} + \lvert J \rvert^{1/n} \leq \left(\prod_{i=1}^n\int q_i(\lambda_i)d\lambda_i\right)^{1/n}.
\end{equation}
Finally, since every $q_i(\lambda_i)$ is a valid probability density function each integral evaluates to unity and we arrive at
\begin{equation}
\lvert I \rvert^{1/n} + \lvert J \rvert^{1/n} \leq 1.
\end{equation}
\begin{flushright}
$\blacksquare$
\end{flushright} 

Conversely, we can also show that the above inequality is tight, by finding an explicit $n$-local decomposition which is able to saturate the bound. As such, consider the following strategies
\begin{align}\label{e:Nlocal strat}
P(a_i|x_i\lambda_i\mu_i) &=
  \begin{cases}
   1 & \text{if } a_i = \lambda_i \oplus \mu_i x_i, \\
   0       & \text{otherwise.} 
  \end{cases}\\
  P(b|y\lambda) &=
  \begin{cases}
   1 & \text{if } b =  \bigoplus_i \lambda_i, \\
   0       & \text{otherwise.} 
  \end{cases}\nonumber
\end{align}
where $q(\lambda_i = 0) = \frac{1}{2}$ and $q(\mu_i = 0) = r$, for all $i$. The $\lambda_i$ are shared variables between each of the Alice's and Bob, whilst the $\mu_i$ are sources of local randomness for each Alice. By direct substitution we see immediately that if $\mu_i = 0$ for all $i$, which occurs with probability $r^n$, then $I = 1$ and $J = 0$. On the other hand, when $\mu_i = 1$ for all $i$, which occurs with probability $(1-r)^n$, then $I = 0$ and $J = 1$. In all other cases we find that $I = 0$ and $J = 0$. Since the values of $I$ and $J$ obtained are simply convex combinations of those obtained for fixed values of the $\mu_i$, the above shows that using the $n$-local strategy \eqref{e:Nlocal strat} we generate points of the form $(I,J) = (r^n,(1-r)^n)$. Such points clearly satisfy the relation $I^{1/n} + J^{1/n} = 1$, i.e. they saturate the inequality. Finally, by considering appropriate symmetries of the above strategy it is easy to generate all possible combinations of signs for $I$ and $J$, hence showing that the above inequality is tight.  

\begin{figure}
\centering
\includegraphics[width=0.9\columnwidth]{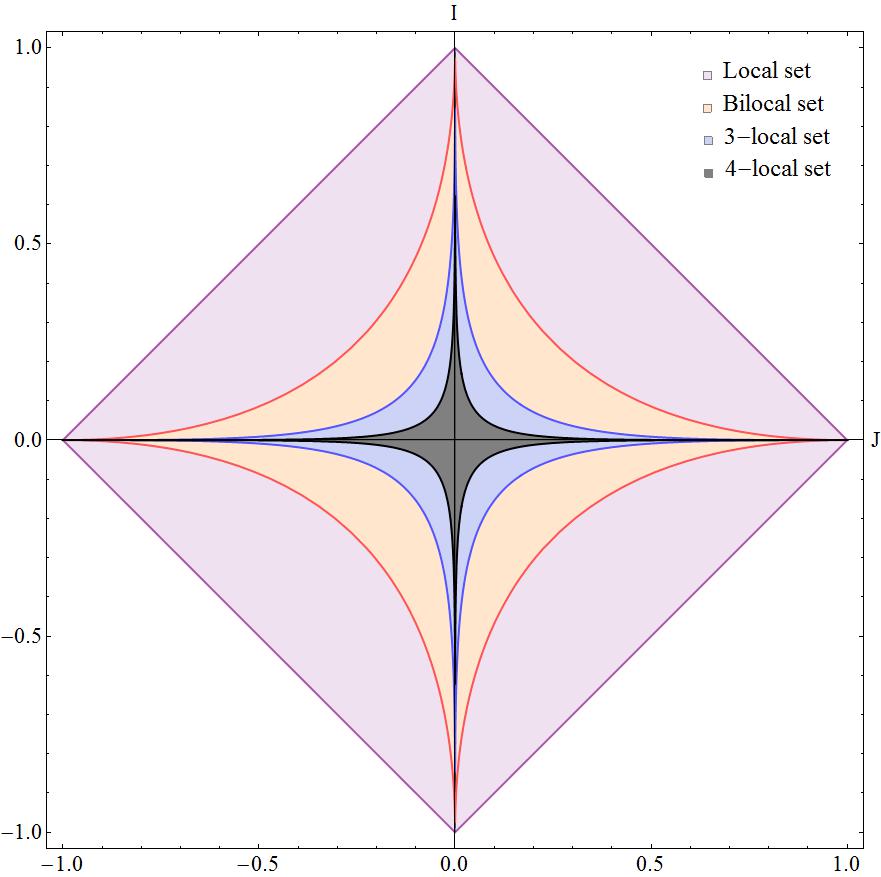} 
\caption{The $n$-local set of probability distributions in the $(I,J)$ plane for the Bob $2 \to 2$ scenario. The rotated square enclosing the $n$-local set is the boundary of the local set.}\label{f:comp}
\end{figure}

We end by noting that the above shows that the vertices, of the form $(I,J) = (\pm 1, 0)$, $(0,\pm 1)$ are obtained from local product distributions, corresponding to no shared randomness between the parties. In the absence of the the $n$-locality constraint one can therefore obtain the convex hull of these points in the $(I,J)$ plane, corresponding to $|I| + |J| \leq 1$. That such points cannot be obtained in an $n$-local manner is because the required shared randomness would need to be shared amongst all $n+1$ parties. In Fig.~\ref{f:comp} we show the show the difference between the $n$-local set and the local set for the case of $n = 2,3,4$, highlighting the difference that these two assumptions make. 

\subsection{Bob $1\rightarrow 2^k$ Scenario}
While the relevance of the previous section is motivated in part by experimental limitations, the analogue $n$-party star-network measurement scenario with the $n$ parties choosing from $2$ measurements and with Bob always performing a fixed measurement on the $n$ qubits at his disposal and obtaining one of $2^n$ possible outcomes, is of greater theoretical and conceptual interest. Bob's measurement will typically be chosen as a complete generalised Bell state measurement since such a measurement allows Bob to perform entanglement swapping to the $n$ distant Alices, which intuition suggests should be difficult to reproduce in a purely $n$-local manner. 

Although Bob obtains $n$ raw bits $\tilde{b}^1\cdots \tilde{b}^{n}$ from his measurement, consider the possibility of generating $k$ bits $b^i = f_i(\tilde{b}^1,\ldots, \tilde{b}^{n})$ from these $n$, and use these processed bits to define all further quantities. The reason to do so will become clearer when we come to study the quantum violations of the inequality which will be presented below.

We start by defining almost the same correlators as in \eqref{n correlator} but with a slight modification necessitated by the change of scenario,
\begin{multline}
\langle A_{x_1}^1 ... A_{x_n}^n B^i  \rangle = \sum_{\substack{a_1\cdots a_n \\ b^i\cdots b^{k}}}  (-1)^{\sum_{j=1}^n a_j + b^i}\\ \times P(a_1\cdots a_n b^1\cdots b^{k}|x_1\cdots x_n),
\end{multline}
where we note that the only difference is in terms of the $b^i$; whereas previously we had the outcome of each measurement for Bob, we now have the bits which comprise the $2^k$ possible outcomes of the single measurement, after the classical processing to generate the bits $b^j$. To proceed we need to define new quantities, which will replace $I$ and $J$. Our method for constructing such quantities relies on the GHZ paradox, and are geared towards quantum violations which we will demonstrate in Section \ref{s: quantum 2 to 2n-1}

Let us define $k = 2^{n-1}$ quantities $I_j$, all of which have the following form:
\begin{equation}\label{new I}
I_j=\frac{1}{2^n} \sum_{x_1\cdots x_n} (-1)^{g_j(x_1,\ldots, x_{n})} \langle A_{x_1}^1 \cdots A_{x_n}^n B^j \rangle,
\end{equation}
each depending on a function $g_j(x_1,\ldots, x_{n})$. For these $2^{n-1}$ functions we take linear functions (with coefficients equal to 1) which contain an even number of $x_i$. For example, the case of $n = 2$ is given by
\begin{align}
g_1(x_1,x_2) &= 0, & g_2(x_1,x_2) &= x_1 + x_2,
\end{align}
(which coincides with the definitions given in \cite{BraRosGis12}), while the case for $n = 3$ is given by
\begin{align}
g_1(x_1,x_2,x_3) &= 0, & g_2(x_1,x_2,x_3) &= x_1 + x_2, \nonumber \\
g_3(x_1,x_2,x_3) &= x_1 + x_3, & g_4(x_1,x_2,x_3) &= x_2 + x_3.
\end{align}
Having defined these quantities, we can now state our second theorem:
\begin{flushleft}
\textbf{Theorem 2 (Bob $1 \rightarrow 2^{2^{n-1}}$ $n$-locality)}
\end{flushleft} Consider the star-network scenario composed by the $n$ edge parties applying two dichotomic measurements, \ie $m_\rma = d_\rma = 2$, and the central party applying a single measurement $m_\rmb = 1$, producing $d_\rmb = 2^{n-1}$ bits after classical processing. In this scenario, every $n$-local probability distribution $\eqref{nlocal p}$ satisfies the following inequality
\begin{equation}\label{e:ineq 1 to 2n}
\sum_{j=1}^{2^{n-1}}\lvert I_j \rvert ^{1/n} \leq 2^{n-2}.
\end{equation}
\textbf{Proof:}\newline\newline
The proof is essentially the same as for theorem 1. Each term $I_j$ can be re-expressed in a form similar to \eqref{e:final I}. Lemma 1 in the Appendix can then be used to arrive at an expression similar to \eqref{e: after lemma 1}. Finally, we note that $2^{n-2}$ functions depend on a given $x_i$, while $2^{n-2}$ are independent of it. As such, a term analogous to the left-hand-side of \eqref{e: upper bound} appears $2^{n-2}$ times, which then gives the bound in a straightforward manner.
\begin{flushright}
$\blacksquare$
\end{flushright} 
It is no longer so straightforward to find an $n$-local strategy which can saturate the above inequality for arbitrary points $(I_1,\ldots, I_{2^{n-1}})$. However, it is possible to show that that in the most relevant direction (i.e. that which we are able to explore with a Bell state measurement and maximally entangled states) the the inequality is tight. This direction corresponds to the symmetric case $I_1 = I_2 = \cdots = I_{2^{n-1}}$. In this case, the strategy
\begin{align}\label{e:Nlocal strat 2k}
P(a_i|x_i\lambda_i\mu_i) &=
  \begin{cases}
   1 & \text{if } a_i = \lambda_i \oplus \mu_i x_i, \\
   0       & \text{otherwise.} 
  \end{cases}\\
  P(b^i|\lambda) &=
  \begin{cases}
   1 & \text{if } b^i =  \bigoplus_i \lambda_i, \\
   0       & \text{otherwise.} 
  \end{cases}\nonumber\\
  q(\lambda_i) = q(\mu_i) &= \tfrac{1}{2}. \nonumber
\end{align}
Considering now a specific choice for the string of local variables $\mu = \mu_1\cdots\mu_n$. If this string contains an even number of ones (i.e. has even parity), then the $I_j$ with $g_j(x_1,\ldots, x_n)$ which contains $x_i$'s only for those $i$ for which $\mu_i = 1$ will evaluate to 1, whilst all other $I_j$ will evaluate to zero, which can easily be verified by inspection. If on the other hand the string of $\mu$ contains an odd number of ones, then all $I_j = 0$, again by inspection. Thus mixing over all strategies (each of which occurs with equal probability $1/2^n$) we can achieve the point $(I_1,\ldots, I_{2^{n-1}}) = (1/2^n,\ldots, 1/2^n)$, such that 
\begin{equation}
\sum_{j=1}^{2^{n-1}} |I_j|^{1/n} = \sum_{j=1}^{2^{n-1}} \tfrac{1}{2} = 2^{n-2},
\end{equation}
which is thus seen to saturate the $n$-locality bound in this direction.
\section{Quantum violations}
Having provided Bell-type inequalities for the star-network measurement scenarios and characterized the $n$-local set in the $2 \rightarrow 2$ and $1\rightarrow 2^n$ settings, we now study the quantum properties of these particular inequality. 

\subsection{Quantum violations in the Bob $2 \rightarrow 2$ scenario}
To show that inequality \eqref{e:ineq 2 to 2} can be violated by quantum mechanics, we need to find a good choice of states to distribute between Alice and Bob, $\rho_{\rma_i \rmb_i}$, as well as choices for the two measurements for each of the Alices, $M_{a_i|x_i}$ and Bob $M_{b|y}$, such that
\begin{multline}
P(a_1\cdots a_n b|x_1\cdots x_n y) \\
= \Tr\Big(M_{a_1|x_1}\otimes\cdots\otimes M_{a_1|x_1}\otimes M_{b|y} \\ \times \rho_{\rma_1 \rmb_1} \otimes \cdots \otimes \rho_{\rma_n \rmb_n}\Big).
\end{multline}
Let us consider that the states distributed correspond to the maximally entangled state $\kets{\psi^-}{\rma_i \rmb_i} = \frac{1}{\sqrt{2}}\left(\ket{0}\ket{1} - \ket{1}\ket{0}\right)_{\rma_i\rmb_i}$. Since the state $\ket{\psi^-}$ is rotationally invariant under $U\otimes U$ operations, it follows that, one the one hand, that the first measurement of each Alice can be taken to be the same without loss of generality, and on the other hand, that the only relevant parameter is the angle between the first fixed measurement and the second, as defined on the Bloch sphere. We will take this angle to be the same in each case, and choose $A^i_0 = \frac{X+Z}{\sqrt{2}}$, $A^i_1 = \frac{X-Z}{\sqrt{2}}$, (i.e. such that $M_{0|x_i} = \tfrac{1}{2}(1+A^i_x)$ and $M_{1|x_i} = \tfrac{1}{2}(1-A^i_x)$) corresponding Alice measures in the $\pm 45^\circ$ in the $xz$ plane, and 

Although the $n$-local set is a non-convex body, thus seemingly making numerical optimisations difficult, as we show in the Appendix, by restricting to a subset of the $n$-local set characterised by the relation $I = \alpha J$, for fixed real $\alpha$, and after having made explicit choices for both the states distributed in the network and the two measurements of Alice, the final optimisation over the measurements of Bob is given by a semi-definite program (SDP), which can then be readily solved using numerical packages such as {\sc cvx} \cite{CVX} for {\sc matlab}. 

In this case, we find that the optimal measurements for Bob are given by parity measurements in the $x$ and $z$ basis. More precisely, the measurement operators $M_{b|y}$ are given by
\begin{align}
M_{b|0} &= \sum_{b_1\oplus\cdots\oplus b_n = b} \Pi^{x}_{b_1}\otimes \cdots \otimes \Pi^{x}_{b_n}, \nonumber\\
M_{b|1} &= \sum_{b_1\oplus\cdots\oplus b_n = b} \Pi^{z}_{b_1}\otimes \cdots \otimes \Pi^{z}_{b_n},
\end{align}
where $\Pi^{x}_{b_i} = \tfrac{1}{2}\left(1+(-1)^{b_i}X\right)$ and $\Pi^{z}_{b_i} = \tfrac{1}{2}\left(1+(-1)^{b_i}Z\right)$ are projectors onto the positive and negative subspaces of $X$ and $Z$ respectively. 

Crucially, such parity measurements can be performed by Bob by simply making the same measurement (either $X$ or $Z$) on each of his $n$ systems, and declaring as outcome the parity of the results obtained from each individual measurement. In other words, his optimal strategy (given the fixed choice of state and measurements for the Alices) here can be viewed as a wiring \cite{Bar07,ShoPopGis06}. It also follows immediately that Bob does not perform any form of entanglement swapping in this scenario, since such separable measurements are incapable of doing so. 

Finally, it is also now straightforward to write down the probability distribution that such a measurement strategy generates, and hence explicitly calculate the violation. Indeed, on each pair $\kets{\psi^-}{\rma_i\rmb_i}$ the measurements performed by Alice and Bob are exactly those which are optimal for violating the CHSH Bell inequality \cite{ClaHorShi69}. The probability distribution that they generate is given by
\begin{equation}
P(a_i b_i | x_i y_i) = \tfrac{1}{4}\left(1 + \tfrac{1}{\sqrt{2}}(-1)^{a_i + b_i + x_i y_i}\right).
\end{equation}
By performing the local wiring strategy of Bob, the final probability distribution thus obtained is
\begin{multline*}
P(a_1\cdots a_n b|x_1\cdots x_n y) = \sum_{b_1 \cdots b_{n-1}}P(a_1b_1|x_1 y)\times \cdots \\
\times P(a_{n-1}b_{n-1}|x_{n-1} y) P(a_n b\oplus b_1\oplus\cdots\oplus b_{n-1}|x_n y),
\end{multline*}
\begin{align}
= &\frac{1}{4^n}\sum_{b_1 \cdots b_{n-1}} \left(1 + \tfrac{1}{\sqrt{2}}(-1)^{a_1 + b_1 + x_1 y}\right) \times \cdots \nonumber \\
&\times \left(1 + \tfrac{1}{\sqrt{2}}(-1)^{a_{n-1} + b_{n-1} + x_{n-1} y}\right),\nonumber \\
&\quad\times \left(1 + \tfrac{1}{\sqrt{2}}(-1)^{a_n + b + b_1 + \cdots + b_{n-1} + x_n y}\right)\nonumber \\
= &\frac{1}{4^n}\sum_{b_1 \cdots b_{n-1}} \Big(1 + \ldots \nonumber \\
&+ \left(\tfrac{1}{\sqrt{2}}\right)^n(-1)^{a_1 + \cdots + a_n + b + y(x_1 + \cdots + x_n)}\Big),
\end{align}
where in the last line the $2^{n-1}$ terms in the sum not written out explicitly contain a factor of the form $(-1)^{b_i + \cdots + b_j}$, containing a subset of the $b_k$. Upon performing the sum, all such terms evaluate to zero, and hence the final expression is given by
\begin{multline}\label{e:opt quantum}
P(a_1\cdots a_n b|x_1\cdots x_n y) = \tfrac{1}{2^{n+1}}\Big(1 + \left(\tfrac{1}{\sqrt{2}}\right)^n\\ \times(-1)^{a_1 + \cdots + a_n + b + y(x_1 + \cdots + x_n)}\Big).
\end{multline}
Interestingly, by comparison with the definitions \eqref{I} and \eqref{J} (using \eqref{n correlator} to express them in full) we see the close relation between the above behaviour and the quantities $I$ and $J$. In particular, we note that when $y = 0$, $P(a_1\cdots a_n b|x_1\cdots x_n y=0)$ contains correlations of the form  $a_1 \oplus \cdots \oplus a_n \oplus b = 0$, exactly those required to obtain a large value of $I$, whilst when $y=1$, $P(a_1\cdots a_n b|x_1\cdots x_n y=1)$ contains correlations of the form  $a_1 \oplus \cdots \oplus a_n \oplus b = x_1 \oplus \cdots \oplus x_n$, required to obtain a large value of $J$. A direct calculation shows that $P(a_1\cdots a_n b|x_1\cdots x_n y)$ from \eqref{e:opt quantum} obtains the values
\begin{align}
I &= \frac{1}{\sqrt{2^n}},& J &= \frac{1}{\sqrt{2^n}},
\end{align}
and thus we obtain
\begin{equation}
|I|^{1/n} + |J|^{1/n} = \sqrt{2} > 1,
\end{equation}
thus demonstrating a quantum violation of the $n$-locality inequality for star-network configurations for all $n$. In particular, assuming only that the the measurements of each Alice are unbiased (unitarily equivalent to measuring $X$ and $Z$), and that a maximally entangled state is distributed on the network, this is the optimal quantum violation achievable. On the other hand, since we saw that this behaviour arises by performing a classical post-processing (wiring) of Bob on a behaviour which violates the CHSH inequality, we see that the violation arises precisely due to the nonlocality of each pair, which contradicts the $n$-locality assumption purely on the basis of violating the assumption of there being a local hidden variable model reproducing the statistics of a single pair. It follows that the critical resistance to white noise and critcal detection efficiency are those of the CHSH inequality, $1/\sqrt{2}$ and $82.8\%$ respectively. Nevertheless, for completeness, and in order to understand exactly how they arise, in the following two subsections we re-derive these two results.

\subsubsection{Resistance to white noise}
Let us assume that the $n$ sources do not produce maximally entangled states, but rather Werner states $\rho_{\rma_i\rmb_i}(v_i)$ with noise parameter $v_i$ of the form
\begin{equation}
\rho_{\rma_i\rmb_i}(v_i) = v_i\ket{\psi^-}\bra{\psi^-}_{\rma_i\rmb_i} + (1-v_i)\frac{\openone}{4}.
\end{equation}
Since it is known that the optimal strategy to violate CHSH with such states is to use the same measurements, which in turn generates the probability distribution
\begin{equation}
P_{v_i}(a_i b_i | x_i y_i) = \tfrac{1}{4}\left(1 + \tfrac{v_i}{\sqrt{2}}(-1)^{a_i + b_i + x_i y_i}\right),
\end{equation}
the final probability distribution generated is 
\begin{multline}\label{e:opt quantum}
P(a_1\cdots a_n b|x_1\cdots x_n y) = \tfrac{1}{2^{n+1}}\Big(1 + \left(\tfrac{1}{\sqrt{2}}\right)^n\textstyle{ \prod_i v_i} \\ \times(-1)^{a_1 + \cdots + a_n + b + y(x_1 + \cdots + x_n)}\Big).
\end{multline}
Denoting by $V =  \prod_i v_i$ the values thus obtained are given by
\begin{align}
I &= \frac{V}{\sqrt{2^n}},& J &= \frac{V}{\sqrt{2^n}},
\end{align}
and
\begin{equation}
|I|^{1/n} + |J|^{1/n} = V^{1/n}\sqrt{2},
\end{equation}
which implies a violation whenever $V > 1/\sqrt{2^n}$. Assuming all sources emit states with the same visibility $v_i = v$, we thus see that each source must emit states with $v > 1/\sqrt{2}$. This is exactly as to be expected, as this is the critical visibility for the Werner state to violate the CHSH inequality, and we saw above that we can precisely view the $n$-locality inequality violation as arising from the violation of CHSH between Bob and each of the Alices. 

\subsubsection{Resistance to detection inefficiency}
For experimental purposes it is interesting to consider the scenario where the involved parties have non-perfect detectors. Since each Alice makes a one-qubit measurement and in the optimal case we have seen that it is sufficient for Bob to make $n$ one-qubit measurements, it is reasonable to assume that all detectors are subject to the same efficiency $\eta$. Using again the insight that each pair is violating the CHSH inequality, we use the optimal binning strategy for this case, which is for both parties to deterministically output 0 whenever they have no click at their detector. 

To see that such a strategy indeed works, we can use the fact that Bobs operators are tensor products, i.e. $B_i = B_i^1 \otimes \cdots \otimes B_i^n$ and that furthermore they act on direct product states, hence correlations factorise, $ \langle B_i^1 \otimes \cdots \otimes B_i^n \rangle = \langle B_i^1\rangle \cdots \langle B_i^n\rangle$. This allows us to re-express $I$ and $J$ as 
\begin{align}\label{e: CHSH form}
I &= \prod_{i=1}^n\tfrac{1}{2}\left(\langle A_0^i B^i_0\rangle + \langle A^i_1 B^i_0\rangle\right), \nonumber \\
J &= \prod_{i=1}^n\tfrac{1}{2}\left(\langle A_0^i B^i_1\rangle - \langle A^i_1 B^i_1\rangle\right) .
\end{align}
Using moreover the symmetry of the strategy (that all Alices measure the same operators), we write $\langle A_0^i B^i_0\rangle = \langle A_0 B_1\rangle$,\footnote{This is a clash of notation, but one which should hopefully cause no confusion given the context.} so that the inequality becomes
\begin{multline}\label{e: CHSH form 2}
|I|^{1/n} + |J|^{1/n} \\= \tfrac{1}{2}|\langle A_0 B_0\rangle + \langle A_1 B_0\rangle| + \tfrac{1}{2}|\langle A_0 B_1\rangle - \langle A_1 B_1\rangle|,
\end{multline}
i.e. in a form which is now manifestly equal to the CHSH inequality. Therefore if we use maximally entangled states $\eta_c = \frac{2}{1+\sqrt{2}} \approx 0.828$, while with partially entangled states we can approach $\eta_c \to \tfrac{2}{3}$ by measuring partially entangled states (in both cases using the optimal strategies for CHSH). 

We note however that there are two alternative possibilities which we leave unexplored. The first is to consider that the parties do not apply a binning strategy, but instead announce the no-click event as a third outcome. This however requires one to move beyond the case of binary outcome $n$-locality inequalities, for which there is currently nothing known, and is an interesting direction for further research. The second possibility is to consider different measurement strategies for Bob. Although it is sufficient for him to make $n$ measurements in the middle, given that his measurement outcome is a single bit one may consider alternative measurement schemes, coherent on all $n$ systems, with the use of only a single detector. Again, we leave the exploration of such alternative strategies, and any potential advantages they may offer, to future work. 

\subsubsection{Exploring the quantum set}
In this final subsection we briefly explore further the structure of the quantum $n$-local set of correlations. In particular, we are interested in the region in the $(I,J)$ plane that we can achieve by performing measurements on the a quantum state, since we know that the entire region specified by inequality \eqref{e:ineq 2 to 2} can be achieved in an $n$-local manner. We will restrict our attention to to the case where the singlet state $\ket{\psi^-}$ is distributed between the parties, and consider that each Alice measurements the same observables $A_0$ and $A_1$ given by
\begin{align}
A_0 &= \cos \theta X + \sin \theta Z,& A_1 &= \cos \phi X + \sin \phi Z.
\end{align}
From equations \eqref{e: CHSH form} and \eqref{e: CHSH form 2} we see that we only need to calculate the 4 correlators that appear in the CHSH inequality. These are given by
\begin{align}
\expect{A_0 B_0} &= -\cos \theta,& \expect{A_0 B_1} &= -\sin \theta, \nonumber \\
\expect{A_1 B_0} &= -\cos  \phi,& \expect{A_1 B_1} &= -\sin \phi, 
\end{align}
from which is follows immediately that we can obtain the values
\begin{align}
I &= \left|\frac{\cos \theta + \cos \phi}{2}\right|^n,& J &= \left|\frac{\sin \theta - \sin \phi}{2}\right|^n.
\end{align}
The optimal is clearly seen to be achieved when $\phi = -\theta$, leading finally to
\begin{align} \label{e: quantum IJ}
I &= \left|\cos \theta\right|^n,& J &= \left|\sin \theta \right|^n.
\end{align}
When $\theta = \tfrac{\pi}{4}$, we recover the previous result of $|I|^{1/n} + |J|^{1/n} = \sqrt{2}$, as should be the case, since in this case the measurements are identical. More generally, we see that \eqref{e: quantum IJ} shows that we can achieve points of the form
\begin{equation}
|I|^{2/n} + |J|^{2/n} \leq 1,
\end{equation}
the boundary being achieved by measurements on the singlet, and thus the interior be measurements on a Werner state. This result shows that we are able to obtain a quadratic-like improvement by using quantum mechanics over classical $n$-local correlations. Finally, we observe that the set of $n/2$-local correlations (for $n$ an even integer) coincides with the set of quantum $n$-local correlations. 

\subsection{Quantum violations in the Bob $2 \rightarrow 2^{2^{n-1}}$ scenario}\label{s: quantum 2 to 2n-1}
Once again, our goal is to demonstrate a good set of measurements and states which is able to violate the inequality \eqref{e:ineq 1 to 2n}. For clarity of presentation we will begin by considering the case $n = 3$ of three Alices. The case for general $n$ is a straightforward generalisation which we will outline afterwards. 

The `raw' measurement we will consider in the middle is a generalised Bell state measurement. Defining the $3$-party GHZ state as
\begin{equation}
\ket{\mathrm{GHZ}_3} = \frac{1}{\sqrt{2}}\left(\ket{0}\ket{0}\ket{0} + \ket{1}\ket{1}\ket{1}\right),
\end{equation}
we can define the generalised Bell basis $\ket{\psi_{i j k}}$ through
\begin{align}
\ket{\psi_{ijk}} &= Z^{i}\otimes X^{j} \otimes X^{k}\ket{\mathrm{GHZ}_3}, \nonumber  \\
&= \frac{1}{\sqrt{2}}\sum_\ell (-1)^{i\ell}\ket{\ell}\ket{\ell\oplus j}\ket{\ell\oplus k}.
\end{align}
We will distribute the state $\kets{\phi^+}{\rma_i \rmb_i} = \frac{1}{\sqrt{2}}\left(\ket{0}\ket{0} + \ket{1}\ket{1}\right)_{\rma_i\rmb_i}$ between the Alice's and Bob, and make use of the following relation
\begin{multline}\label{e: max entangled max entangled}
\kets{\phi^+}{\rma_1 \rmb_1}\kets{\phi^+}{\rma_2 \rmb_2}\kets{\phi^+}{\rma_3 \rmb_3} \\= \frac{1}{\sqrt{2^3}}\sum_{ijk}\kets{\psi_{ijk}}{\rma_1\rma_2\rma_3}\kets{\psi_{ijk}}{\rmb_1\rmb_2\rmb_3}.
\end{multline} 
That is, we can alternatively see the tensor product of $\ket{\phi^+}$ states distributed in the star-network configuration as a maximally entangled state between the $2^3$ generalised Bell states of Alice and of Bob. This relation shows that upon performing a such a generalised Bell state measurement in the middle entanglement is swapped to the $3$ edge Alices. 

As mentioned previously, the construction of the $n$-locality inequality was designed specifically with the GHZ paradox in mind. Namely, we have the following relations
\begin{align}
X\otimes X \otimes X \ket{\mathrm{GHZ}_3} &= \ket{\mathrm{GHZ}_3}, \nonumber \\
Y\otimes Y \otimes X \ket{\mathrm{GHZ}_3} &= -\ket{\mathrm{GHZ}_3},\nonumber \\
Y\otimes X \otimes Y \ket{\mathrm{GHZ}_3} &= -\ket{\mathrm{GHZ}_3}, \\
X\otimes Y \otimes Y \ket{\mathrm{GHZ}_3} &= -\ket{\mathrm{GHZ}_3}. \nonumber 
\end{align}
Using the relations $XZ^{i} = (-1)^{i}Z^{i}X$ and $YX^{i} = (-1)^{i}X^{i}Y$ we see that the Bell basis $\ket{\psi_{ijk}}$ therefore satisfies
\begin{align}\label{e: stab bell}
X\otimes X \otimes X \ket{\psi_{ijk}} &= (-1)^{i}\ket{\psi_{ijk}}, \nonumber \\
Y\otimes Y \otimes X \ket{\psi_{ijk}} &= (-1)^{i+j+1}\ket{\psi_{ijk}}, \nonumber \\
Y\otimes X \otimes Y \ket{\psi_{ijk}} &= (-1)^{i+k+1}\ket{\psi_{ijk}}, \\
X\otimes Y \otimes Y \ket{\psi_{ijk}} &= (-1)^{i+j+k+1}\ket{\psi_{ijk}}.\nonumber 
\end{align}
In the above $ijk = \tilde{b}^1\tilde{b}^2\tilde{b}^3$ are the raw outcomes of Bob's Bell state measurement. Motivated by the above relations, we define the bits $b^1 b^2 b^3 b^4$ as
\begin{align}
b^1 &= i, & b^2 &= i \oplus j \oplus 1, \nonumber \\
b^3 &= i \oplus k \oplus 1, & b^4 &= i \oplus j \oplus k \oplus 1,
\end{align}
which are seen to correspond to the 4 exponents in \eqref{e: stab bell}. Considering only a single one of these bits at a time, we can think of the raw Bell state measurement as being a `parent' measurement for 4 effective measurements
\begin{align}
M^\ell_{b_\ell} = \sum_{ijk} b^\ell \ket{\psi_{ijk}}\bra{\psi_{ijk}},
\end{align}
such that $M^\ell_{0} + M^\ell_{1} = \openone_\rmb$. Upon performing such a measurement, Bob projects the Alices onto the (unnormalised) state
\begin{align}\label{e: steering}
&\Tr_\rmb \Bigg(\tfrac{1}{2^3}\sum_{\substack{ijk\\i'j'k'}} \left(\ket{\psi_{ijk}}\bra{\psi_{i'j'k'}}\otimes\ket{\psi_{ijk}}\bra{\psi_{i'j'k'}}\right)\openone_\rma \otimes M_{b_\ell}^\ell\Bigg), \nonumber \\
&= \tfrac{1}{2^3}\sum_{ijk} b_\ell \ket{\psi_{ijk}}\bra{\psi_{ijk}} = \tfrac{1}{2^3}M_{b_\ell}^\ell.
\end{align}
Now, by construction the $M^\ell_{b_\ell}$ have the property that all terms appearing in $M^\ell_{0}$ satisfy the $\ell^{\mathrm{th}}$ stabiliser relation in \eqref{e: stab bell} with sign $+1$ while $M^\ell_{1}$ satisfies it with sign $-1$, for example 
\begin{align}\label{e: XXX example}
(X\otimes X \otimes X) M^0_0 &= M^0_0 & (X\otimes X \otimes X) M^0_1 &= -M^0_1.
\end{align}
Thus, the above shows that each $M^\ell_{b_\ell}$ lies in the $(-1)^{\ell}$ eigenspace of the corresponding stabilizer string. Since this space is 4 dimensional, and $M^\ell_{b_\ell}$ is a sum of 4 orthogonal projectors, we see that the operators constructed are in fact equal to the projector onto the respective eigenspace, and so we have the relation that 
\begin{equation}\label{e:stab relation}
S_\ell = M^\ell_{0} - M^\ell_{1}, 
\end{equation}
where $S_\ell$ is the $\ell^\mathrm{th}$ stabilizer string from \eqref{e: stab bell}.

Crucially, this shows that, conditioned on $b^\ell$, if the Alices were able to measure the observable $S_\ell$  then they would find correlations between the joint result of all of their measurements and the outcome of the measurement of Bob. 

Finally, we note that the Alices can gain information about all of the $S_\ell$ simultaneously, not by measuring $X$ and $Y$ directly, but instead by measuring in the rotated bases $A_0^i = \frac{X+Y}{\sqrt{2}}$, $A_1^i = \frac{X-Y}{\sqrt{2}}$, since the following two obvious identities hold
\begin{align}\label{e: identity}
\tfrac{X}{\sqrt{2}} &= \tfrac{1}{2}\left(\tfrac{X+Y}{\sqrt{2}} + \tfrac{X-Y}{\sqrt{2}}\right)& \tfrac{Y}{\sqrt{2}} &= \tfrac{1}{2}\left(\tfrac{X+Y}{\sqrt{2}} - \tfrac{X-Y}{\sqrt{2}}\right).
\end{align}
Indeed, consider the quantity $I_1$, with $g_1(x_1,x_2,x_3) = 0$. The quantum value can be written, denoting $\rho = \ket{\phi^+}\bra{\phi^+}_{\rma_1\rmb_1}\otimes\ket{\phi^+}\bra{\phi^+}_{\rma_2\rmb_2}\otimes\ket{\phi^+}\bra{\phi^+}_{\rma_3\rmb_3}$ as
\begin{align}
I_1&=\frac{1}{2^3} \sum_{x_1x_2x_3} \langle A_{x_1}^1 A_{x_2}^2 A_{x_3}^3 B^1 \rangle, \nonumber \\
&= \frac{1}{2^3} \sum_{x_1x_2x_3}\Tr\Big(\left(A_{x_1}^1 \otimes A_{x_2}^2 \otimes A_{x_3}^3 \otimes (M_0^1-M_1^1)\right)\rho\Big),\nonumber \\
&= \frac{1}{16\sqrt{2}}\Tr\Big(\left(X \otimes X \otimes X \right)\left(X \otimes X \otimes X \right)\Big), \nonumber \\
&= \frac{1}{16\sqrt{2}} \Tr\left(\openone \otimes \openone \otimes \openone \right)= \frac{1}{2\sqrt{2}},
\end{align}
where to go from the second line to the third line we used the identities \eqref{e: identity} to evaluate the sums, \eqref{e: steering} to eliminate Bob, and \eqref{e:stab relation} to re-express $M_0^1-M_1^1$ in terms of $S_1$

The same analysis can be applied to the quantities $I_2$, $I_3$ and $I_4$, which correspond to the the $2^{\mathrm{nd}}$, $3^{\mathrm{rd}}$ and $4^{\mathrm{th}}$ relation in \eqref{e: stab bell} respectively. In all three cases, it is straightforward to see that $I_j = \frac{1}{2\sqrt{2}}$ is achieved using the above strategy. Putting everything together then, we find
\begin{equation}
\sum_{j=1}^4 |I_j|^{1/3} = 4\times \tfrac{1}{\sqrt{2}} = 2\sqrt{2} > 2,
\end{equation}
thus demonstrating a quantum violation in the Bob $1 \to 2^4$ scenario with 3 Alices. 

The generalisation to the case of $n$ Alices is straightforward given the above. The generalised Bell basis extends to the $n$ party basis $\ket{\psi_{i_1\cdots i_n}}$ straightforwardly, as
\begin{align}
\ket{\psi_{i_1\cdots i_n}} &= Z^{i_1}\otimes X^{i_2} \otimes \cdots \otimes X^{i_n}\ket{\mathrm{GHZ}_n}, \nonumber  \\
&= \frac{1}{\sqrt{2}}\sum_\ell (-1)^{i_1\ell}\ket{\ell}\ket{\ell\oplus i_2}\cdots\ket{\ell\oplus i_n},
\end{align}
the analogue of \eqref{e: max entangled max entangled} holds, 
\begin{multline}\label{e: gen max entangled max entangled}
\kets{\phi^+}{\rma_1 \rmb_1}\cdots\kets{\phi^+}{\rma_n \rmb_n} \\= \frac{1}{\sqrt{2^n}}\sum_{i_1\cdots i_n}\kets{\psi_{i_1\cdots i_n}}{\rma_1\cdots \rma_n}\kets{\psi_{i_1\cdots i_n}}{B_1\cdots B_n},
\end{multline} 
and stabilizer relations analogous to \eqref{e: stab bell} hold, with the set containing all $2^{n-1}$ strings of operators which contain an even number of $Y$ operators. To each of these relations corresponds a quantity $I_j$ as defined in \eqref{new I} and a measurement operator $M_{b_\ell}^\ell$, 
\begin{align}
M^\ell_{b_\ell} = \sum_{i_1\cdots i_b} b^\ell(i_1,\cdots, i_n) \ket{\psi_{i_1 \cdots i_n}}\bra{\psi_{i_1 \cdots i_n}},
\end{align}
for functions $b^\ell(i_1,\cdots, i_n)$ which following immediately from the property the Bell basis with respect to each stabiliser relation, for which the property $M^\ell_{0} - M^\ell_{1} = S_\ell$ holds. Upon performing such measurements, the unnormalised states that Bob prepares are now given by
\begin{align}\label{e: steering gen}
&\Tr_\rmb\! \Bigg(\!\tfrac{1}{2^3}\!\!\sum_{\substack{i_1\cdots i_n\\i_1'\cdots i_n'}}\!\! \left(\ket{\psi_{i_1\cdots i_n}}\bra{\psi_{i_1'\cdots i_n'}}\otimes\ket{\psi_{i_1\cdots i_n}}\bra{\psi_{i_1'\cdots i_n'}}\right)\!\openone_\rma \!\otimes\! M_{b_\ell}^\ell\Bigg), \nonumber \\
&= \tfrac{1}{2^n}\sum_{i_1\cdots i_n} b_\ell \ket{\psi_{i_1\cdots i_n}}\bra{\psi_{i_1\cdots i_n}} = \tfrac{1}{2^n}M_{b_\ell}^\ell.
\end{align}
All together, denoting now $\rho = \ket{\phi^+}\bra{\phi^+}_{\rma_1\rmb_1}\otimes\cdots\otimes\ket{\phi^+}\bra{\phi^+}_{\rma_n\rmb_n}$ we find
\begin{align}
I_j&=\frac{1}{2^n} \sum_{x_1\cdots x_n} (-1)^{g_j(x_1,\ldots, x_n)}\langle A_{x_1}^1 \cdots A_{x_n}^n B^j \rangle, \nonumber \\
&= \frac{1}{2^n} \sum_{x_1\cdots x_n}\Tr\Big(\left(A_{x_1}^1 \otimes \cdots \otimes A_{x_n}^n \otimes (M_0^j-M_1^j)\right)\rho\Big),\nonumber \\
&= \frac{1}{2^n\sqrt{2^n}}\Tr\Big(S_jS_j\Big), \nonumber \\
&= \frac{1}{2^n\sqrt{2^n}} \Tr\openone^n= \frac{1}{\sqrt{2^n}}.
\end{align}
Since the above holds for all $j$, we obtain the value
\begin{equation}
\sum_{j=1}^{2^{n-1}} |I_j|^{1/n} = 2^{n-1}\times \tfrac{1}{\sqrt{2}} = 2^{n-2}\sqrt{2} > 2^{n-2},
\end{equation}
demonstrating that for all $n$ we obtain a quantum $n$-local violation which is $\sqrt{2}$ times larger than the classically obtainable value. 

\subsubsection{Resistance to white noise} 
To calculate the resistance to white noise in this scenario, we make use of the following relation
\begin{multline}
\Tr_\rmb\Big(\left(\rho_{\rma_1\rmb_1}(v_1)\otimes \cdots \otimes \rho_{\rma_n\rmb_n}(v_n)\right)\left(M_0^\ell-M_1^\ell\right) \Big) \\
= \frac{1}{2^n}\prod_i v_i \left(M_0^\ell-M_1^\ell\right),
\end{multline}
from which we immediately calculate, again denoting $V = \prod_i v_i$,
\begin{align}
I_j=& \frac{1}{2^n} \sum_{x_1\cdots x_n}\Tr\Big(\big(A_{x_1}^1 \otimes \cdots \otimes A_{x_n}^n \otimes (M_0^j-M_1^j)\big),\nonumber \\
& \times \left(\rho_{\rma_1\rmb_1}(v_1)\otimes \cdots \otimes \rho_{\rma_n\rmb_n}(v_n)\right)\Big),\nonumber \\
&= \frac{V}{2^n\sqrt{2^n}}\Tr\Big(S_jS_j\Big), \nonumber \\
&= \frac{V}{2^n\sqrt{2^n}} \Tr\openone^n= \frac{V}{\sqrt{2^n}}.
\end{align}
Therefore we find
\begin{equation}
\sum_{j=1}^{2^{n-1}} |I_j|^{1/n} = 2^{n-1}\times \tfrac{V^{1/n}}{\sqrt{2}} = V^{1/n}2^{n-2}\sqrt{2} ,
\end{equation}
and violation is obtained whenever $V^{1/n}2^{n-2}\sqrt{2} > 2^{n-2}$, i.e. when $V> 1/\sqrt{2^n}$, the same robustness to noise as obtained in the Bob $2\to 2$ $n$-locality scenario.

In summary, we therefore see that by extending from the bilocality to the $n$-locality scenario, with the inequalities presented here, no advantage has yet been found over the standard Bell nonlocality scenario for demonstrating the quantum nature of maximally entangled bipartite states. However, from \cite{CavAlmSca11} it is already known that distributing quantum states in the star-network configuration with sufficiently many parties and measurements provide a stronger test of nonlocality, even without invoking the $n$-locality assumption (i.e. even when all parties are allowed to share local hidden variables). It thus remains an open problem to find better inequalities which are more robust to white noise for maximally entangled states, by choosing $m_\rma > 2$, i.e. by using more measurements for the edge parties.

\subsubsection{Resistance to detection inefficiency}
Unlike in the Bob $2\to 2$ scenario, where Bob could be taken to be making $n$ single party measurements, thus allowing us to study the case where all parties had inefficient detectors, here the fact that Bob is making a single Bell state measurement complicates the situation. Indeed, in linear optical set-ups it is not even possible to perform such a measurement, demonstrating that any analysis here needs to be tailored to a specific implementations. To avoid such complications we will focus only on the simpler scenario where Bob is taken to perform an ideal measurement $\eta_\rmb = 1$, while all Alice's will be taken to use inefficient detectors with $\eta_{\rma_i} = \eta$ for each Alice.  

In this simple situation it is however straightforward to see that all strategies available to the Alices for producing an outcome when receiving a no click perform equally well. This is because of the fact that the inequality \eqref{e:ineq 1 to 2n} is an \emph{full correlation} $n$-locality inequality, \ie the value of the inequality explicitly only depends upon joint correlations among all parties. Now, in any instance when only a subset of parties have a no-click event there can never be joint correlations between all parties, and hence they necessarily achieve the value 0 for the inequality. The only instance where one could potentially win is when all parties have a no click event, in which case a deterministic strategy can again produce correlations amongst all parties. However, since we now take Bob to have unit efficiency detectors such events never occur, and therefore all strategies perform equally. 

Furthermore, from the above it is clear that the value obtained is
\begin{equation}
\sum_{j=1}^{2^{n-1}}\lvert I_j \rvert ^{1/n} = 2^{n-1}\times \tfrac{\eta}{\sqrt{2}} = 2^{n-2}\eta\sqrt{2}, 
\end{equation}
and therefore $\eta_c = 1/\sqrt{2}$. 

As in the previous case of Bob $2 \to 2$ we leave it for future research to extend this analysis to more complicated situations involving detector inefficiency also for Bob. 

\section{Conclusion}
In this work we have studied a generalisation of the bilocality scenario to one involving $n$ edge parties distributed in a star-network. We have shown how the previous technique of deriving inequalities satisfied by all local correlations can be naturally extended to derive new inequalities which are structurally very similar to those previously known. We showed furthermore that quantum mechanics is able to violate these inequalities for all values of $n$, and that the violation can be understood intuitively through the use of the GHZ paradox for the case of a full generalised Bell state measurement.

Unfortunately, we found that neither the critical robustness to white noise nor the critical detection efficiency are lowered for maximally entangled states, relative to simply performing a standard CHSH Bell test. This is disappointing since the power of distributing quantum states in a star-network configuration was already demonstrated in \cite{CavAlmSca11} where it was shown that one can lower the critical robustness to noise even in the absence of the  $n$-locality assumption (i.e. even if all $n+1$ nodes jointly shared hidden variables the behaviour could not be explained). This suggests that by combining the star-network with the $n$-locality assumption should lead to stronger tests, but that to do so will probably require going to larger input sizes for the edge parties. However, as it currently stands there is still no known situation where $n$-locality offers an advantage over simply performing a standard Bell test.

Our work also suggests a number of other possibilities for future work. In the first instance, we have considered here only a single geometry of the parties, whereas one can consider many inequivalent geometries, with sources distributed amongst various groups of parties. This ties in the with line of research recently initiated on causal networks \cite{Fri12,Fri14,HenLalPus14}, and the approach of deriving new inequalities would complement this line of research. Initial results in this direction will be presented in \cite{Tav14}. A second direction is to further study the derivation of $n$-locality inequalities based upon the stabilizer formalism. Indeed, the construction presented here based upon the GHZ paradox is a particular instance of this, and it would be interested to see if inequalities can be derived based upon arbitrary graph states. This may in turn help to develop the theory of $n$-locality for larger numbers of inputs and outcomes, the natural place now to look for advantages over Bell nonlocality.

\textbf{Acknowledgements} We thank N. Brunner, J. Bowles, J. Bohr Brask, M. T\'{u}lio Quintino and D. Roset  for discussions. This work was supported by the EU ERC CoG QITBOX, the John Templeton Foundation, and through Marie Curie COFUND actions through a Beatriu de Pin\'os fellowship (BP-DGR 2013) and an ICFOnest fellowship.

\begin{appendix}

\section{Lemma 1}
Here we prove lemma 1. \newline\newline
\textbf{Lemma 1:} Let $x_i^k$ be non-negative real numbers and $m,n\in\mathbb{N}$, then
\begin{equation}\label{e: lemma 1}
\sum_{k=1}^{m}\left(\prod_{i=1}^{n} x_i^k\right)^{1/n} \leq \prod_{i=1}^{n} \left(x_i^1+x_i^2+...+x_i^m\right)^{1/n}.
\end{equation}
\textbf{Proof:} \newline\newline 
We make use of the elementary fact that the arithmetic mean is always larger than or equal to the geometric mean of a sequence. Exploiting this fact, we can make $m$ inequalities on the form
\begin{equation}
\prod_{i=1}^n\left(\frac{x_i^l}{x_i^1+x_i^2+...+x_i^m}\right)^{1/n} \leq \frac{1}{n} \sum_{i=1}^{n} \frac{x_i^l}{x_i^1+x_i^2+...+x_i^m},
\end{equation}
for $l=1,2,...,m$. Now, by taking the sum of the left and right hand sides, over the $m$ inequalities
\begin{align}\label{e:lemma 1 2} 
\sum_{l=1}^{m} \prod_{i=1}^n\left(\frac{x_i^l}{x_i^1+x_i^2+...+x_i^m}\right)^{1/n} \nonumber \\
\leq \frac{1}{n} \sum_{i=1}^{n} \sum_{l=1}^{m}\frac{x_i^l}{x_i^1+x_i^2+...+x_i^m}, \nonumber \\ 
=\frac{1}{n}\sum_{i=1}^{n} \frac{\sum_{k=1}^{m}x_i^l}{x_i^1+x_i^2+...+x_i^m},\nonumber  \\
=\frac{n}{n}=1.
\end{align} 
Finally, multiplication of both sides of \eqref{e:lemma 1 2} with $\prod_{i=1}^{n}\left(x_i^1+x_i^2+...+x_i^m\right)^{1/n}$ yields \eqref{e: lemma 1}.

\section{Obtaining partial results using Semi-Definite-Programming}
In this appendix we will show that although the set of $n$-local correlations is not a convex set, by making appropriate restrictions to a subset of the correlations, one obtains a convex set which is allows for the use of Semi-Definite-Programming (SDP) techniques to perform optimisations.

The general problem that one would like to solve, is to find a set of quantum states $\kets{\psi}{\rma_i\rmb_i}$ distributed between the Alices and Bob in the star-network configuration, along with a set of measurements $M_{a_i|x_i}$ for each Alice and measurement $M_{b_1\cdots b_k}$ or measurements $M_{b|y}$, depending on the scenario ($1 \to 2^k$, and $2\to 2$ respectively), which maximise the the quantities appearing in \eqref{e:ineq 2 to 2} or \eqref{e:ineq 1 to 2n}, of the form
\begin{equation}
S = |I|^{1/n} + |J|^{1/n}.
\end{equation}
Although these inequalities are non-linear, it is possible to instead perform a linear optimisation by restricting to a subset of correlations. In particular, since in both instances the quantities $I$ and $J$ are linear functions of the behaviours $P(a_0\cdots a_n b_1\cdots b_k|x_0\cdots x_1)$ and $P(a_0\cdots a_n b|x_0\cdots x_1 y)$ respectively, by restricting to the subset defined by
\begin{equation}\label{e: constraint}
I = \alpha J,
\end{equation}
for $\alpha$ a fixed real constant, by maximising $I$ alone, subject to the linear constraint \eqref{e: constraint}, we obtain also the maximal value of $S$ in this subset, equal to $S_\alpha = I^{1/n}\left(1 + 1/\alpha^{1/n}\right)$. 

Furthermore, by making an explicit choice for the the distributed states $\kets{\psi}{\rma_i \rmb_i}$, the natural choice being pure maximally or partially entangled states, as well as choices for the measurements of each Alice $M_{a_i|x_i}$, the remaining optimisation over the measurements of Bob, either $M_{b_1\cdots b_k}$ or $M_{b|y}$, depending on the context, reduces to an SDP. Let us focus on the Bob $2\to 2$ case, with the Bob $1 \to 2^k$ case being almost identical. To see the reduction to an SDP, let us consider how to re-express $I$, namely 
\begin{widetext}
\begin{align}
I &= \frac{1}{2^n} \sum_{x_1,\ldots, x_n} \langle A_{x_1}^1 \cdots A_{x_n}^n B_0 \rangle, \nonumber \\
&= \frac{1}{2^n} \sum_{\substack{x_1\cdots x_n\\a_1\cdots a_n b}}(-1)^{b+\sum_i a_i} P(a_1\cdots a_n b|x_1\cdots x_n y=0), \nonumber \\
&= \frac{1}{2^n} \sum_{\substack{x_1\cdots x_n\\a_1\cdots a_n b}}(-1)^{b+\sum_i a_i} \Tr\left(\left(M_{a_1|x_1}\otimes\cdots\otimes M_{a_1|x_1}\otimes M_{b|0}\right) \left(\rho_{\rma_1 \rmb_1} \otimes \cdots \otimes \rho_{\rma_n \rmb_n}\right)\right), \nonumber \\
&= \frac{1}{2^n} \sum_{\substack{x_1\cdots x_n\\a_1\cdots a_n b}}(-1)^{b+\sum_i a_i} \Tr_B\Bigg(\Big(\Tr_{\rma_1}\big((M_{a_1|x_1}\otimes \openone_{B_1})\rho_{\rma_1 \rmb_1}\big)\otimes\cdots\otimes \Tr_{\rma_n}\big((M_{a_n|x_n}\otimes \openone_{B_n})\rho_{\rma_n \rmb_n}\big)\Big)M_{b|0}\Bigg).\end{align}
\end{widetext}
Defining operators
\begin{equation}
\Lambda_i = \sum_{x_ia_i}(-1)^{a_i}\Tr_{\rma_i}\big((M_{a_i|x_i}\otimes \openone_{B_i})\rho_{\rma_i \rmb_i}\big),
\end{equation}
which are fixed, given a fixed choice of state and measurements for each Alice, we can finally express $I$ as
\begin{equation}
I = \frac{1}{2^n} \sum_b (-1)^b \Tr_{B} \big(\left(\Lambda_1\otimes\cdots\otimes\Lambda_n\right)M_{b|0}\big),
\end{equation}
which is of the form $\sum_b \Tr( C_b^\dagger M_{b|0})$, i.e. a linear function of the $M_{b|0}$. By introducing furthermore the fixed operators
\begin{equation}
\Omega_i = \sum_{x_ia_i}(-1)^{a_i+x_i}\Tr_{\rma_i}\big((M_{a_i|x_i}\otimes \openone_{B_i})\rho_{\rma_i \rmb_i}\big),
\end{equation}
an identical analysis shows that we can write
\begin{equation}
J = \frac{1}{2^n} \sum_b (-1)^b \Tr_{B} \big(\left(\Omega_1\otimes\cdots\otimes\Omega_n\right)M_{b|1}\big)
\end{equation}
Thus the optimisation problem which we wish to solve is given by the following SDP
\begin{align}
I^* &= \max_{M_{b|y}} \quad\frac{1}{2^n} \sum_b (-1)^b \Tr_{B} \big(\left(\Lambda_1\otimes\cdots\otimes\Lambda_n\right)M_{b|0}\big),\nonumber \\
\mathrm{s.~t.}& \quad\sum_b (-1)^b \Tr_{B} \big(\left(\Lambda_1\otimes\cdots\otimes\Lambda_n\right)M_{b|0}\big), \nonumber \\
& = \alpha \sum_b (-1)^b \Tr_{B} \big(\left(\Omega_1\otimes\cdots\otimes\Omega_n\right)M_{b|1}\big), \\
& M_{b|y} \geq 0 \quad \forall b,y \quad\quad\quad \sum_b M_{b|y} = \openone \quad \forall y.\nonumber
\end{align}
Finally, for the case of Bob $1\to 2^k$, an identical analysis to above shows that the optimisation problem is instead given by the following SDP
\begin{align}
I^* &= \max_{M_{b_1\cdots b_k}} \frac{1}{2^n} \!\!\sum_{b_1\cdots b_k}\!\! (-1)^{b_1} \Tr_{B} \big(\left(\Lambda_1\otimes\cdots\otimes\Lambda_n\right)M_{b_1\cdots b_k}\big),\nonumber \\
\mathrm{s.~t.}& \sum_{b_1\cdots b_k}\!\! (-1)^{b_1} \Tr_{B} \big(\left(\Lambda_1\otimes\cdots\otimes\Lambda_n\right)M_{b_1\cdots b_k}\big),  \\
& = \alpha \sum_{b_1\cdots b_k} (-1)^{b_2} \Tr_{B} \big(\left(\Omega_1\otimes\cdots\otimes\Omega_n\right)M_{b_1\cdots b_k}\big), \nonumber \\
& M_{b_1\cdots b_k} \geq 0 \quad \forall b_1\cdots b_k \quad\quad\quad \sum_{b_1\cdots b_k} M_{b_1\cdots b_k} = \openone \nonumber.
\end{align}
\end{appendix}

\end{document}